# Improvement of the Orthogonal Code Convolution Capabilities Using FPGA Implementation

Naima Kaabouch, *Member, IEEE*, Aparna Dhirde, *Member, IEEE*, Saleh Faruque, *Member, IEEE*
Department of Electrical Engineering, University of North Dakota, Grand Forks, ND 58202-7165
University of North Dakota, Grand Forks, ND 58202-7165

*Abstract* - **When data is stored, compressed, or communicated through a media such as cable or air, sources of noise and other parameters such as EMI, crosstalk, and distance can considerably affect the reliability of these data. Error detection and correction techniques are therefore required. Orthogonal Code is one of the codes that can detect errors and correct corrupted data. An n-bit orthogonal code has n/2 1s and n/2 0s. In a previous work these properties have been exploited to detect and correct errors. In this paper we present a new methodology to enhance error detection capabilities of the orthogonal code. The technique was implemented experimentally using Field Programmable Gate Arrays (FPGA). The results show that the proposed technique improves the detection capabilities of the orthogonal code by approximately 50%, resulting in 99.9% error detection, and corrects as predicted up to (n/4-1) bits of error in the received impaired code with bandwidth efficiency.**

## 1. Introduction

When data is stored, compressed, or communicated through a media such as cable or air, sources of noise and other parameters such as EMI, crosstalk, and distance can considerably affect the reliability of these data. Error detection and correction techniques are therefore required. Some of those techniques can only detect errors, such as the Cyclic Redundancy Check [1, 2]; others are designed to detect as well as correct errors, such as Salomon Codes [3]. However, the existing techniques are not able to achieve high efficiency and to meet bandwidth requirements, especially with the increase in the quantity of data transmitted. Orthogonal Code is one of the codes that can detect errors and correct corrupted data.

Our objective in this paper is to enhance the error control capabilities of orthogonal codes by means of Field Programmable Gate Array (FPGA) implementation.

## 2. Orthogonal codes

Orthogonal codes are binary valued and they have equal number of 1's and 0's. An n-bit orthogonal code has n/2 1's and n/2 0's; i.e., there are n/2 positions where 1's and 0's differ [4, 5]. Therefore, all orthogonal codes will generate zero parity bits. The concept is illustrated by means of an 8-bit orthogonal code as shown in Fig.1. It has 8-orthogonal codes and 8-antipodal codes for a total of 16-biorthogonal codes. Antipodal codes are just the inverse of orthogonal codes; they are also orthogonal among themselves.

| Orthogonal Code | | | | | | | | | | Antipodal Code | | | | | | | |
|---|---|---|---|---|---|---|---|---|---|---|---|---|---|---|---|---|---|
| 0 | 0 | 0 | 0 | 0 | 0 | 0 | 0 | | | 1 | 1 | 1 | 1 | 1 | 1 | 1 | 0 |
| 0 | 1 | 0 | 1 | 0 | 1 | 0 | 1 | 0 | | 1 | 0 | 1 | 0 | 1 | 0 | 1 | 0 | 0 |
| 0 | 0 | 1 | 1 | 0 | 0 | 1 | 1 | 0 | | 1 | 1 | 0 | 0 | 1 | 1 | 0 | 0 | 0 |
| 0 | 1 | 1 | 0 | 0 | 1 | 1 | 0 | 0 | | 1 | 0 | 0 | 1 | 1 | 0 | 0 | 1 | 0 |
| 0 | 0 | 0 | 0 | 1 | 1 | 1 | 1 | 0 | | 1 | 1 | 1 | 1 | 0 | 0 | 0 | 0 | 0 |
| 0 | 1 | 0 | 1 | 1 | 0 | 1 | 0 | 0 | | 1 | 0 | 1 | 0 | 0 | 1 | 0 | 1 | 0 |
| 0 | 0 | 1 | 1 | 1 | 1 | 0 | 0 | 0 | | 1 | 1 | 0 | 0 | 0 | 0 | 1 | 1 | 0 |
| 0 | 1 | 1 | 0 | 1 | 0 | 0 | 1 | 0 | | 1 | 0 | 0 | 1 | 0 | 1 | 1 | 0 | 0 |
| | | | | | | | | | | | | | | | | | |
| 0 | 0 | 0 | 0 | 0 | 0 | 0 | 0 | P | | 0 | 0 | 0 | 0 | 0 | 0 | 0 | 0 | P |

Fig. 1. Illustrations of the proposed concept. An 8-bit orthogonal code has 8 orthogonal codes and 8-antipodal codes for a total of 16 bi-orthogonal codes. All orthogonal and antipodal codes generate zero parity.

Since there is an equal number of 1's and 0's, each orthogonal code will generate a zero parity bit. Therefore, each antipodal code will also generate a zero parity bit. A notable distinction in this method is that the transmitter does not have to send the parity bit since the parity bit is known to be always zero [7]. Therefore, if there is a transmission error, the receiver will be able to detect it by generating a parity bit at the receiving end.

Before transmission a k-bit data set is mapped into a unique n-bit. For example, a 4-bit data set is represented by a unique 8-bit orthogonal code which is transmitted without the parity bit.

When received, the data are decoded based on code correlation. It can be done by setting a threshold midway between two orthogonal codes. This is given by the following equation

$$d_{th} = \frac{n}{4} \qquad (1)$$

Where n is the code length and $d_{th}$ is the threshold, which is midway between two orthogonal codes. Therefore, for the 8-bit orthogonal code (Fig. 2), we have $d_{th} = 8/4 = 2$.

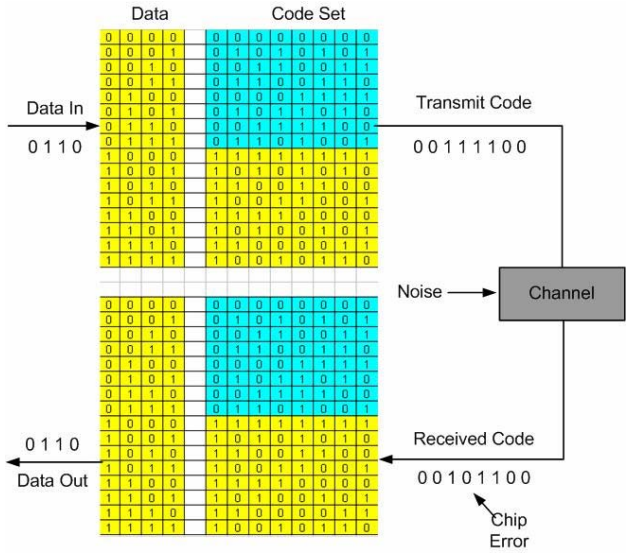

Fig. 2. Illustration of Encoding and Decoding.

This mechanism offers a decision process, where the incoming impaired orthogonal code is examined for correlation with the neighboring codes for a possible match. The acceptance criterion for a valid code is that an n-bit comparison must yield a good auto-correlation value; otherwise, a false detection will occur. This is governed by the following correlation process, where a pair of n-bit codes $x_1, x_2 ... x_n$ and $y_1, y_2 ... y_n$ is compared to yield,

$$R(x,y) = \sum_{i=1}^{n} x_i y_i \leq \frac{n}{4} - 1 \quad (2)$$

Where R(x, y) is the auto correlation function, n is the code length, $d_{th}$ is the threshold defined in (1). Since the threshold ($d_{th}$) is in between two valid codes, an additional 1-bit offset is added to (2) for reliable detection. The average number of errors that can be corrected by means of this process can be estimated by combining (1) and (2), yielding,

$$t = n - R(x,y) = \frac{n}{4} - 1 \quad (3)$$

In (3), t is the number of errors that can be corrected by means of an n-bit orthogonal code. For example, a single error-correcting orthogonal code can be constructed by means of an 8-bit orthogonal code (n = 8). Similarly, a three-error-correcting orthogonal code can be constructed by means of a 16-bit orthogonal code (n = 16), and so on. Table-1 below shows a few orthogonal codes and the corresponding error-correcting capabilities:

TABLE I

Orthogonal Codes and the Corresponding Chip Error Control Capabilities.

| n | t |
|---|---|
| 8 | 1 |
| 16 | 3 |
| 32 | 7 |
| 64 | 15 |

### 3. METHODOLOGY AND FPGA IMPLEMENTATION

#### 3.1 Design Methodology

Since there is an equal number of 1's and 0's, each orthogonal code will generate a zero parity bit. If the data has been corrupted during the transmission the receiver can detect errors by generating the parity bit for the received code and if it is not zero then the data is corrupted. However the parity bit doesn't change for an even number of errors, hence the receiver can only detect errors $2^n/2$ combinations of the received code. Therefore detection percentage is 50% [6]. Our approach is not to use the parity generation method to detect the errors, but a simple technique based on the comparison between the received code and all the orthogonal code combinations stored in a look up table. The technique which involves a transmitter and receiver is described below.

#### 3.2 Transmitter

The transmitter includes two blocks: an encoder and a shift register. The encoder encodes a k-bit data set to $n=2^{k-1}$ bits of the orthogonal code and the shift register transforms this code to a serial data in order to be transmitted as shown in Fig.3. For example, 4-bit data is encoded to 8-bit ($2^3$) orthogonal code according to the lookup table shown in Fig.2. The generated orthogonal code is then transmitted serially using a shift register with the rising edge of the clock.

#### 3.3 Receiver

The received code is processed through the sequential steps, as shown in Fig.4. The incoming serial bits are converted into n-bit parallel codes. The received code is compared with all the codes in the lookup table for error detection. This is done by counting the number of ones in the signal resulting from 'XOR' operation between the received code and each combination of the orthogonal codes in the lookup table. A counter is used to count the number of ones in the resulting n-bit signal and also searches for the minimum count. However a value rather than zero shows an error in the received code. The orthogonal code in the lookup table which is associated with the minimum count is the closest match for the corrupted received code. The matched orthogonal code in the lookup table is the corrected code, which is then decoded to k-bit data. The receiver is able to correct up to (n/4)-1 bits in the received impaired code. However, if the minimum count is associated with more than one combination of orthogonal code then a signal, REQ, goes high.

## 4. IMPLEMENTATION AND RESULTS

A Spartan-3 hardware board and ISE Xillinx software have been used for code testing. The simulation has been performed using ModelSim XE software. The simulation results were verified for most of the combinations of 8-bit and some of the 16-bit orthogonal code. The software simulation results along with the clock cycles are explained for the transmitter and receiver in the following sections.

### 4.1 Transmitter

Fig.5 shows an example of the results of the transmitter simulation corresponding to the input data value "0110" labeled as 'data'. This data has been encoded to the associated orthogonal code "00111100" labeled 'ortho'. The signal 'EN' is used to enable the transmission of the serial bits 'txcode' of the orthogonal code with every rising edge of the clock.

### 4.2 Receiver

Upon reception, the incoming serial data is converted into 8-bit parallel code 'rxcode'. Counter is used to count the number of 1's after XOR operation between the received code and all combinations of orthogonal code in the lookup table. 'Count' gives the minimum count of ones among them. The orthogonal code 'ortho' associated with the minimum count is the closest match for the received code, which is then decoded to the final data given by signal 'data'. Three different cases result from this simulation. In the first case, the received code has a match in the lookup table. As shown in Fig.6, the received code is rxcode= "00111100", count='0' and hence the received code is not corrupted. The code is then decoded to the corresponding final data "0110".

In the second case, the received code has no match in the lookup table. As shown in Fig.7, the received code is rxcode="00110100", the value of minimum count is '1', which reveals an error. The corresponding orthogonal code is ortho= "00111100" which is the closest match for the received code given by the minimum count, and the decoded final data is "0110". In this case the single bit error is detected and corrected by the receiver.

In the third case, there is more than one possibility of closest match in the lookup table. As shown in Fig.8, the received code is rxcode= "00110000". The value of minimum count is associated with more than one orthogonal code and thus it is not possible to determine the closest match in the lookup table for the received code. Then the signal labeled 'REQ' goes high, which is a request for a retransmission.

### 4.3 Results

The results of the simulation show that for a k-bit data, the corresponding n-bit orthogonal code is able to detect any faulty combination other than the combinations of orthogonal code in the lookup table. The numbers of these combinations are $2^k$. Hence the percentage of detection is given by $(2^n - 2^k)/2^n$ %. Similarly, the system is able to correct up to $(n/4)-1$ bit error and the number of clock cycles required to process the received code are $(2n+2)$. For example when a 4-bit data is encoded in to 8-bit orthogonal code; it has $2^4 = 16$ combinations of orthogonal code. Therefore, out of 256 possible combinations of 8-bit received code the receiver will not able to detect error in those codes which are one of the combinations of orthogonal code. Hence the detection percentage for 8-bit orthogonal code is given by $(2^8 - 2^4)/2^8 = 93.57\%$ and also able to correct single bit error. Similarly, the percentage of detection for 16-bit orthogonal code is 99.95% and gives 3-bit of error correction.

Table II shows a summary of the results and their corresponding detection rates for 8-bit, 16-bit, and N-bit orthogonal codes.

TABLE II

Detection rate computed from the results corresponding to 8-bit, and 16-bit orthogonal codes.

|  | Number of combinations | $N_f$ Number of undetected combinations | Detection Rate |
|---|---|---|---|
| **8-bit Codes** | 256 | $N_f$=16 | 93.57 % |
| **16-bit Codes** | 65535 | $N_f$=32 | 99.95% |
| **N-bit Code** | $2^N$ | $N_f$=2N | $(2^N-N_f)/2^N$ |

## 5. CONCLUSION

The results of the orthogonal code implementation show that this technique improved the error detection from 50% to 93% for 8-bit orthogonal code and 99.9% for 16-bit orthogonal code. The technique proposed can be applied to any encoding system used for digital transmission. Future work includes improvement of correction capasbilities of the orthogonal code and paralell implementation to speed up the data processing.

ACKNOWLEDGMENT

This work was supported by the ND EPSCoR project through National Science Foundation grant # UND0012168.

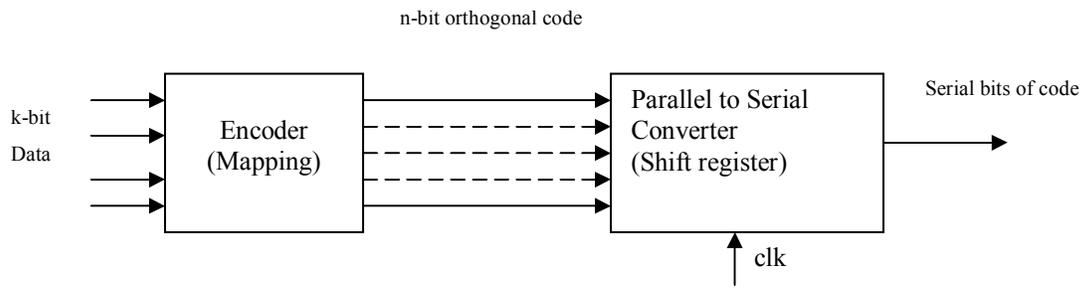

Fig. 3. Block diagram of the transmitter.

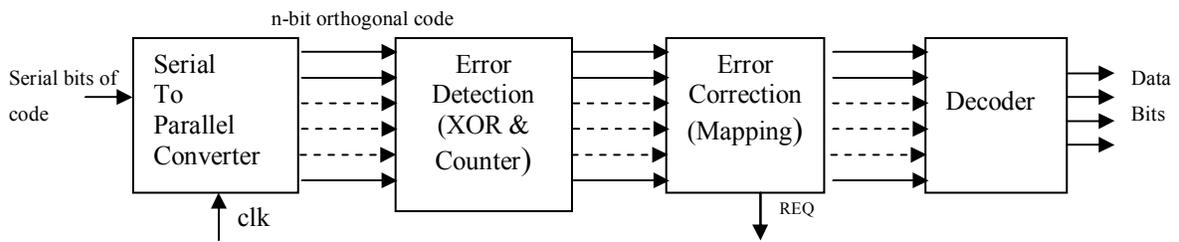

Fig. 4. Block diagram of the receiver.

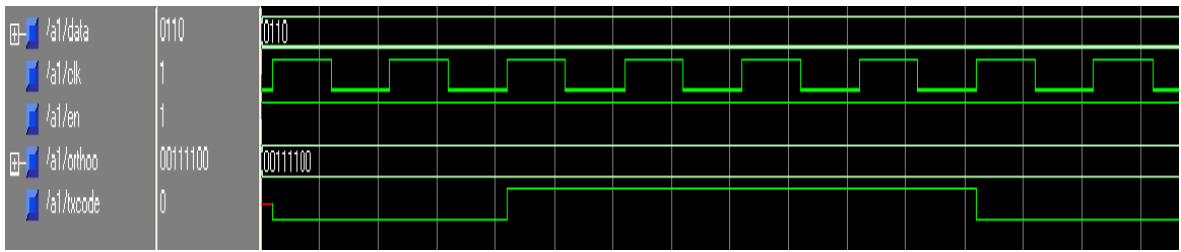

Fig. 5. Example of the simulation results of the transmitter.

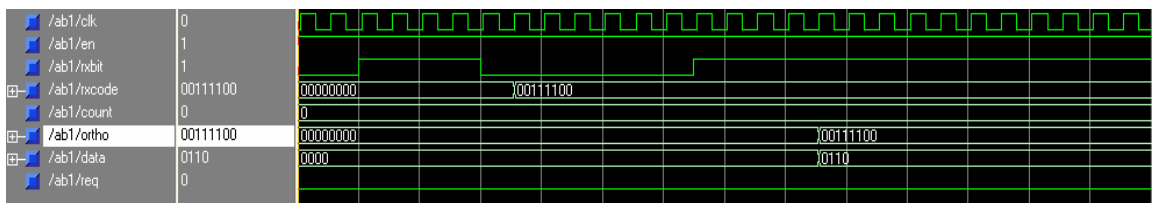

Fig. 6. Example of case one simulation.

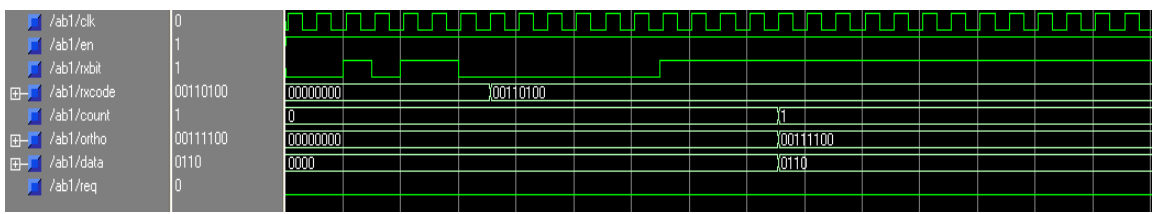

Fig. 7. Example of case two simulation.

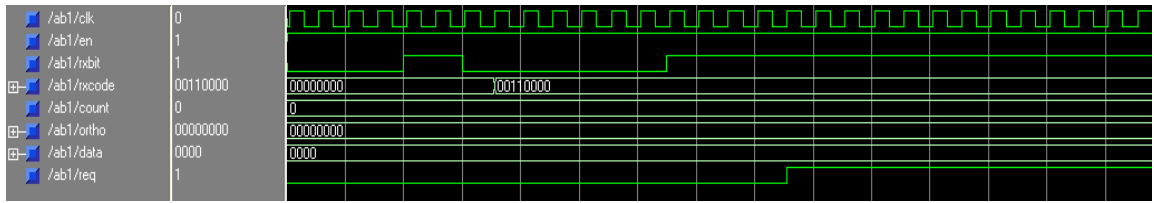

Fig. 8. Example of case three simulation.